\documentclass[acmsmall]{acmart}
\usepackage{siunitx}  
\usepackage{algorithmic}
\usepackage{algorithm}

\newcommand{\e}{\mathrm{e}}

\setcopyright{acmcopyright}
\copyrightyear{0000}
\acmYear{0000}
\acmDOI{XXXXXXX.XXXXXXX}

\acmJournal{TECS}
\acmVolume{00}
\acmNumber{0}
\acmArticle{000}
\acmMonth{0}

\acmSubmissionID{64}

\begin{document}

\title{Online Training and Inference System on Edge FPGA Using Delayed Feedback Reservoir}

\author{Sosei Ikeda}
\affiliation{%
  \institution{Kyoto University}
  \city{Kyoto}
  \country{Japan}}
\email{sikeda@easter.kuee.kyoto-u.ac.jp}

\author{Hiromitsu Awano}
\affiliation{%
  \institution{Kyoto University}
  \city{Kyoto}
  \country{Japan}} 
\email{awano@i.kyoto-u.ac.jp}

\author{Takashi Sato}
\affiliation{%
  \institution{Kyoto University}
  \city{Kyoto}
  \country{Japan}}
\email{takashi@i.kyoto-u.ac.jp}

\renewcommand{\shortauthors}{Ikeda et al.}

\begin{abstract}
%
A delayed feedback reservoir (DFR) is a hardware-friendly reservoir computing system.
Implementing DFRs in embedded hardware requires efficient online training.
However, two main challenges prevent this: hyperparameter selection,
which is typically done by offline grid search, and training of the output linear layer, which is memory-intensive.
This paper introduces a fast and accurate parameter optimization method for
the reservoir layer utilizing backpropagation and gradient descent by adopting a modular DFR model.
A truncated backpropagation strategy is proposed to reduce memory consumption associated with the expansion of the recursive structure while maintaining accuracy.
The computation time is significantly reduced compared to grid search.
Additionally, an in-place Ridge regression for the output layer via 1-D Cholesky decomposition is presented, reducing memory usage to be 1/4.
These methods enable the realization of an online edge training and inference system of DFR on an FPGA,
reducing computation time by about 1/13 and power consumption by about 1/27 compared to software implementation on the same board.
\end{abstract}



\received{00 January 0000}
\received[revised]{00 January 0000}
\received[accepted]{00 January 0000}
\maketitle


\section{Introduction}\label{sec:intro}

Reservoir computing (RC) is a machine learning method that utilizes a reservoir to
nonlinearly transform inputs into a high-dimensional space~\cite{tanaka2019recent}.
Typically, the reservoir weights remain fixed, with only the weights of the output layer being subjected to training~\cite{lukovsevivcius2009reservoir}. 
A delayed feedback reservoir (DFR)~\cite{appeltant2011information} is a specific RC variant, 
which incorporates a reservoir composed of a nonlinear element and a feedback loop.
The nonlinear element commonly adopts a Mackey--Glass equation~\cite{mackey1977oscillation}.
Various implementations of RC have been demonstrated, leveraging electronic circuits~\cite{appeltant2011information}, optical elements~\cite{larger2012photonic}, etc.
Among RC architectures, the structural simplicity of DFR makes it an attractive candidate for embedded hardware implementation due to its low power consumption~\cite{tanaka2019recent}.
In addition, unlike most RC implementations, the reservoir weights in DFR can be adjusted through several parameters, resulting in superior accuracy compared to other RC architectures~\cite{ikeda2022hardware}.

In the context of embedded applications, the concept of an online training and inference system on edge devices is of paramount importance.
The edge system offers several benefits, including real-time capabilities, adaptability, and security~\cite{cao2020overview}.
By executing computation close to the data source, the system curtails data transmission and achieves swift response times.
The system can be fine-tuned to suit the edge environment without necessitating prior offline training.
Additionally, data processing can transpire without external data transmission, thereby minimizing exposure to potential threats.
A common use case for such a system is predictive maintenance of factory equipment.
Traditional preventive maintenance, conducted at regular intervals, often results in excessive or unnecessary upkeep.
Ideally, potential equipment failures should be predicted by analyzing equipment data in real-time.
The online edge system meets this need by providing real-time processing capabilities and specialized equipment fault estimation while concurrently ensuring the security of sensitive equipment.

To implement such a system using DFR, efficient optimization methods for reservoir weights and memory efficient linear regression methods are imperative.
First of all, the time-intensive reservoir weight optimization impedes its online execution.
The performance of RC heavily depends on reservoir weights or parameters~\cite{tanaka2019recent}.
Typically, parameter optimization in DFR relies on grid search~\cite{appeltant2011information},
which confronts notable scalability challenges as the dimension of the parameter space increases.
In DFR, both the regularization parameter of the linear regression at the output layer
and the reservoir parameters necessitate simultaneous optimization.
The objective function within the parameter space can exhibit substantial nonlinearity, mandating finely segmented grids and 
resulting in significant computational overhead.
Additionally, the memory-intensive linear regression at the output layer obstructs its edge implementation.
The number of features provided by the reservoir layer depends on the length of the input series.
Consequently, the features must be converted into a reservoir representation with a fixed length~\cite{ikeda2022hardware}.
The reservoir representation is a high-dimensional vector, approximately $1,000$ in dimension,
as it must encapsulate features from preceding time periods.
Given that the reservoir representation acts as the input for the output layer, deriving its parameters entails computationally expensive operations,
including the inversion of a matrix of corresponding size, such as 1,000 by 1,000.
Storing this high-dimensional matrix and its inverse necessitates a considerable amount of memory.

This paper presents the application of backpropagation and gradient descent to the online optimization of the reservoir layer of DFR\@.
Although backpropagation and gradient descent~\cite{amari1993backpropagation} are widely used and proven effective training methods for neural networks,
to the best of the authors' knowledge, no prior study has applied them to DFR.
There exist two challenges: the use of nonlinear element that are often based on physical phenomena that have no derivatives, and the recursive structure
that requires unfolding and requires infinitely long backpropagation traces~\cite{nakajima2022physical}.
To address these challenges, the authors propose to
leverage the concept of modular DFR~\cite{ikeda2023modular}.
The modular DFR model reduces the number of parameters in the reservoir layer and divides the nonlinear element into multiple blocks, making backpropagation easier to apply.
In addition to using this model, the recursively defined reservoir state of the DFR, which serves as a feature for the output layer, is approximated.
The authors demonstrate that this approach enables fast online training for DFRs.

The authors propose a hardware-friendly Cholesky decomposition for optimizing the output layer of edge DFR 
by proving that the matrix associated with Ridge regression, the most widely used and high-performance learning method, is symmetric and positive definite.
We also present an in-place inversion procedure that uses a single 1-D array to save memory usage.
This is achieved by carefully designing the update order of the matrix in the Cholesky decomposition and matrix inversion, allowing
the original matrix and computed results to share the same array.
In addition, introducing small buffer reduces memory access collisions during the sum-of-products operation and facilitates parallel processing, thereby increasing the operating frequency.

The major contributions of this work are summarized as follows:
\begin{enumerate}
 \setlength{\itemsep}{0pt}
 \item Online edge training and inference system of DFR with a novel and efficient
	parameter optimization for reservoir and output layers.
       The implementation on an FPGA demonstrated significant performance gain by approximately 1/13 and power consumption by about 1/27 compared to the software implementation.
 \item Development of a fast and accurate parameter optimization method utilizing backpropagation and gradient descent by
       adopting a modular DFR model.
       The proposed method achieves accuracy comparable to traditional grid search while reducing computation time by 700x. 
 \item Proposal of an in-place Ridge regression for the output layer via 1-D
       Cholesky decomposition.
       The hardware implementation of this method reduces memory usage to 1/4 while maintaining the same accuracy as the traditional Gaussian elimination method.

\end{enumerate}

The remainder of this paper is organized as follows.
First, Section~\ref{sec:dfr} reviews the general concept of DFR and the modular DFR model.
Section~\ref{sec:propose} proposes the online end-to-end edge DFR system.
As the key components of this system, a fast parameter optimization method and a memory-saving Ridge regression method are proposed.
Section~\ref{sec:evaluation} presents the evaluation of the proposed calculation method and the online edge implementation of DFR\@.
Finally, Section~\ref{sec:conclusion} concludes the paper.


\section{DFR}\label{sec:dfr}

\begin{table}
 \centering
 \caption{Symbol table}
 \label{tab:symbol}
 \begin{tabular}{c|l} \toprule
  Symbol  & Definition \\ \midrule
  $u(k)$ & digital time series input \\
  $i(t)$ & analog signal with D-A conversion of $u(k)$ \\
  $m(t)$ & masking signal \\
  $j(t)$ & masked signal of $i(t)$ \\
  $x(t)$ & output of the nonlinear element \\
  $\gamma$ & input scaling \\
  $\eta$ & parameter of the nonlinear element \\
  $f$ & nonlinear function \\
  $\theta$ & time interval of virtual nodes \\
  $\tau$ & total delay of a feedback loop \\
  $N_x$ & number of virtual nodes; reservoir size \\
  $T$ & length of a time series \\
  $\boldsymbol{r}$ & reservoir representation \\
  $N_r$ & number of features in a reservoir representation \\
  $\boldsymbol{y}$ & output \\
  $N_y$ & number of classes \\
  $W_{\mathrm{out}}$ & weight of output layer \\
  $\boldsymbol{b}$ & bias of output layer \\
  $p$ & parameter of modular DFR \\
  $q$ & parameter of modular DFR \\
  $\boldsymbol{e}$ & one-hot encoding of $\boldsymbol{y}$ \\
  $\beta$ & regularization parameter \\
  $s$ & $N_x^2+N_x+1$ \\
  $L$ & loss function \\ \bottomrule
 \end{tabular}
\end{table}

First, the concept and basic operation of DFR are explained.
Next, the dot-product reservoir representation (DPRR)~\cite{ikeda2022hardware}, which enhances the accuracy of classification problems.
Subsequently, we present the modular DFR model, which allows for more flexible design of nonlinear elements and incorporates
backpropagation and gradient descent, offering an efficient alternative to time-consuming grid search.
Finally, commonly used, memory-intensive output layer training methods are described.
The symbols used throughout the discussion are listed in Table~\ref{tab:symbol}.

\subsection{Concept of analog DFR}\label{subsec:dfr}

Reservoir computing is a machine learning scheme consisting of three layers:
the input layer, the reservoir layer, and the output layer~\cite{lukovsevivcius2009reservoir}.
The input layer converts the input signal into a format suitable for entering the reservoir layer.
The reservoir layer then transforms the signal into a high-dimensional nonlinear vector.
Finally, the output layer performs pattern recognition using a simple learning algorithm.
While the weights of the input and reservoir layers are fixed, the weights of the output layer are determined through training.

The DFR is a specific implementation of RC that incorporates a nonlinear element and a delayed feedback loop.
Its structure captures past inputs, making it particularly suitable for processing time series data~\cite{tanaka2019recent}.
Due to its simplicity, DFR is considered more suitable for hardware implementation than other RC schemes~\cite{tanaka2019recent}.
Typically, hardware realizations of DFRs employ analog circuits in the reservoir layer~\cite{appeltant2011information,soriano2014delay}.

\begin{figure}[tb]
 \centering
 \includegraphics[width=1.0\linewidth]{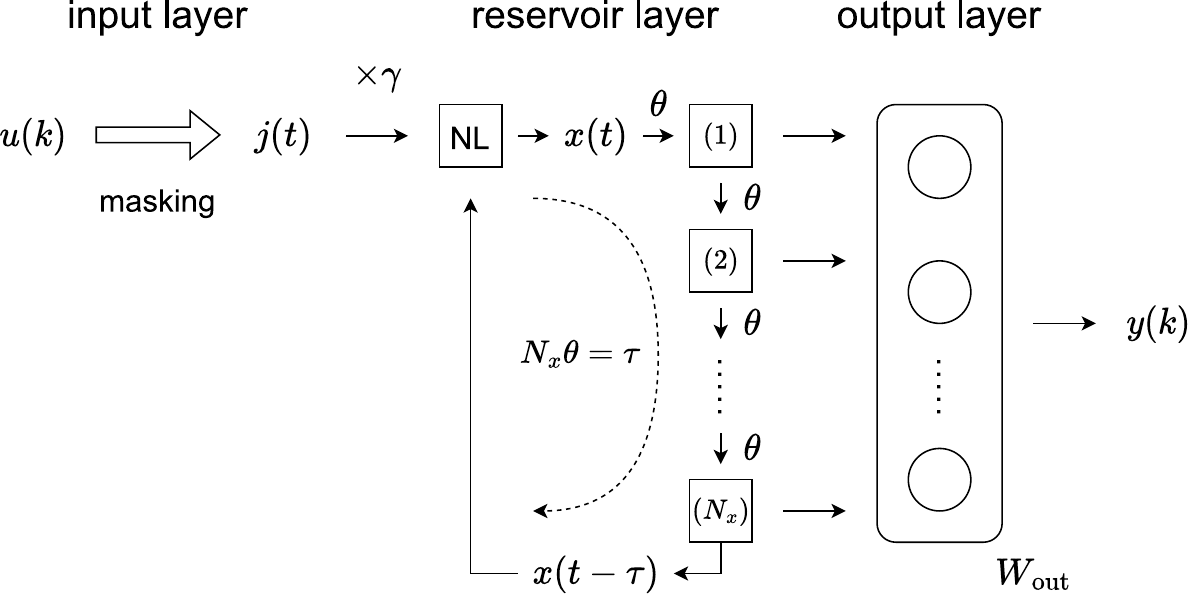}
 \caption{Conceptual diagram of DFR\@.
 The reservoir consists of a NL and a feedback loop with a total delay $\tau$.
 The feedback loop comprises $N_x$ virtual nodes with a time interval $\theta$.
 }\label{fig:dfr}
\end{figure}

To demonstrate the operation and challenges of DFR, we begin with the most standard analog implementation.
Fig.~\ref{fig:dfr} illustrates a conceptual schematic of the DFR\@.
The digital time-series input, $u(k)$, sampled at a period $\tau$, is first converted into an analog signal, $i(t)$.
This signal then undergoes masking, where it is multiplied by the mask signal with a faster sample rate $\theta$.
\begin{figure}[tb]
 \centering
 \includegraphics[width=1.0\linewidth]{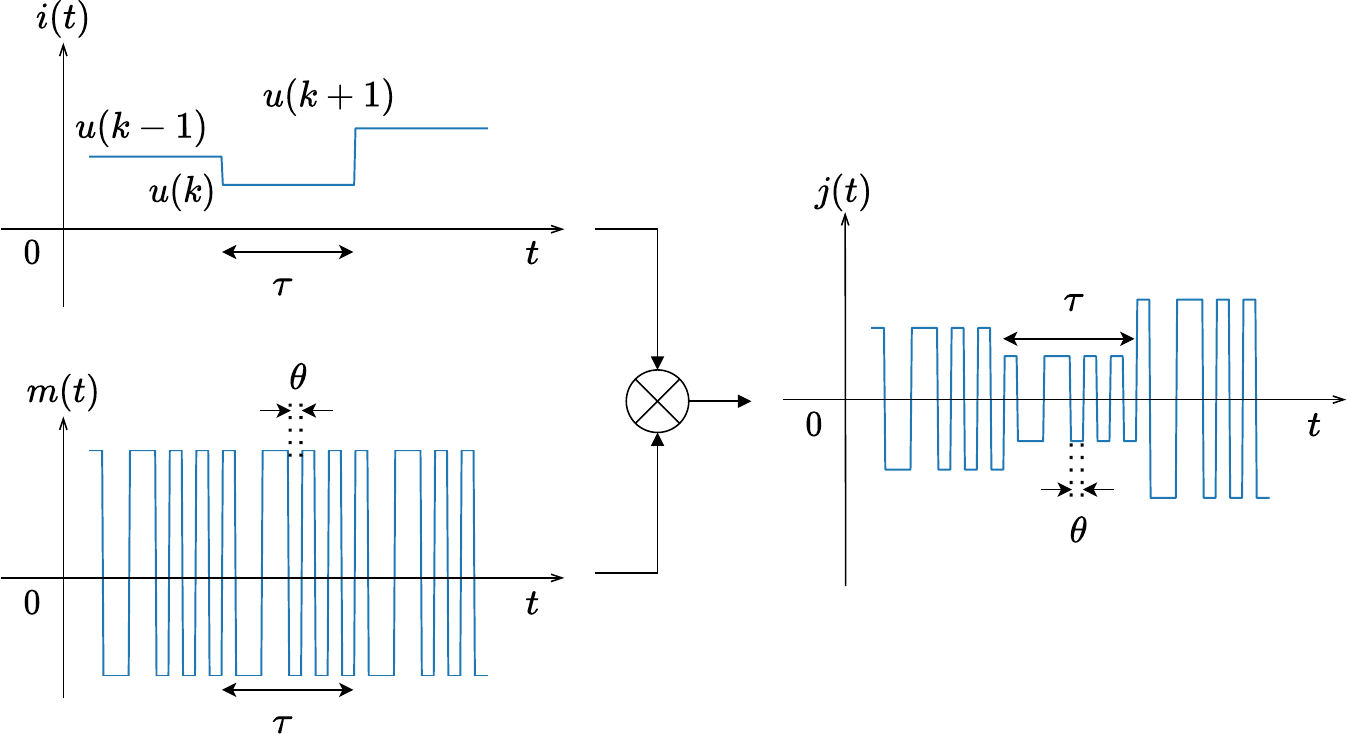}
 \caption{Masking process. $i$ is obtained by digital-to-analog conversion of input signal $u$.
 Signal $i$ is constant at all $\tau$.
 Mask signal $m$ takes different values at all time intervals $\theta$, and its period is $\tau$.
 Input signal to reservoir is expressed as $j(t) = i(t) \cdot m(t)$.}\label{fig:mask}
\end{figure}
Fig.~\ref{fig:mask} depicts the most frequently used mask signals generated by a pseudo-random bit sequence.
However, other masks may also be employed~\cite{appeltant2011information,appeltant2014constructing}.
The resulting signal $j(t)$ is then scaled by $\gamma$, combined with the reservoir feedback, and passed through
a nonlinear element (NL) to produce the output, $x(t)$.

The NL is mathematically represented by the following delay differential equation:
\begin{align}
 \frac{\mathrm d}{{\mathrm d}t}x(t)=F(x(t-\tau), \gamma j(t)).
 \label{eq:delay}
\end{align}
The Mackey--Glass model~\cite{mackey1977oscillation} is a commonly used NL in various studies~\cite{appeltant2011information,soriano2014delay,alomar2015digital}.
Its operation is expressed by the following equations:
\begin{align}
 \frac{\mathrm d}{{\mathrm d}t}x(t)=-x(t)+\eta f(x(t-\tau),\gamma j(t)), \label{eq:MG} \\
 f(x(t-\tau),\gamma j(t))=\frac{[x(t-\tau)+\gamma j(t)]}{1+{[x(t-\tau)+\gamma j(t)]}^p}. \label{eq:f}
\end{align}
Here, $p$ is an adjustable parameter.

A feedback loop with a total delay of $\tau$ comprises virtual nodes spaced at regular time intervals of $\theta$.
The signal values at these virtual nodes serve as the features and are collected as a vector
of $N_x$ elements, referred to as the reservoir state, where $N_x\theta =\tau$.
The number of virtual nodes $N_x$ is also referred to as the reservoir size.

The reservoir state is defined as follows:
\begin{align}
 \boldsymbol{x}(k)\equiv[x(k\tau -\theta ), x(k\tau -2\theta ), \ldots , x(k\tau -\tau)]. \label{eq:x}
\end{align}
The output, representing the inference result, is obtained by applying a linear transformation of the output layer to the reservoir state.
The detailed methods for this process are described in Sections~\ref{subsec:dprr} and \ref{subsec:gaussian}.

\subsection{Concept of digital DFR}\label{subsec:digitalDFR}

Fully digital versions of DFRs have been proposed to ease implementation~\cite{appeltant2011information,alomar2015digital}.
In these digital DFRs, the signal, $\boldsymbol{j}(k)$ after the masking process is expressed as $\boldsymbol{j}(k)=\boldsymbol{m}u(k)$
where $\boldsymbol{m}$ is a mask vector of length $N_x$
and its elements are determined randomly and digitally~\cite{appeltant2011information}.

After masking, the nonlinear element and feedback loop of the DFR determine the reservoir state, $\boldsymbol{x}(k)$.
Assuming that $f$ in Eq.~(\ref{eq:MG}) is constant for a short time $\theta$, which is the time interval between the virtual nodes,
the differential equation in Eq.~(\ref{eq:MG}) is solved for $x(t)$ as follows:
\begin{align}
 x(t)=x_0\e^{-t}+\eta (1-\e^{-t})f(x(t-\tau)+\gamma j(t)),
\end{align}
where $x_0$ is the initial value of $x(t)$ at each $\theta$~\cite{appeltant2011information}.
The value of the next virtual node can be expressed as $x(\theta)$.
Assuming that the components of $\boldsymbol{x}(k)$ and $\boldsymbol{j}(k)$ are
\begin{align}
 \boldsymbol{x}(k)&\equiv[x(k)_1,x(k)_2,\ldots,x(k)_{N_x}],\\
 \boldsymbol{j}(k)&\equiv[j(k)_1,j(k)_2,\ldots,j(k)_{N_x}],
\end{align}
$\boldsymbol{x}(k)$ is obtained recurrently as follows:
\begin{align}
  x(k)_1 &= x(k-1)_{N_x}\e^{-\theta}\nonumber\\
 &+(1-\e^{-\theta})f(x(k-1)_1, j(k)_1), \label{eq:digitalDFR1}\\
  x(k)_n &= x(k)_{n-1}\e^{-\theta}\nonumber\\
 &+(1-\e^{-\theta})f(x(k-1)_n, j(k)_n).\ (n\geq2) \label{eq:digitalDFR2}
\end{align}
The reservoir state is initialized to zero.
The construction and operation of the output layer remain identical to those in the analog implementation.

In both implementations, grid search is the only effective method for parameter optimization
due to the recursive nature of the network construction and the differentially defined NL\@.

\subsection{DPRR}\label{subsec:dprr}

In classification tasks involving time-series inputs, a single output is generated at the end.
To ensure consistency, the features obtained by the reservoir layer should be converted to an intermediate representation with a fixed length,
known as the {\em reservoir representation}~\cite{ikeda2022hardware}.
This conversion enables the features to be processed by an output layer of a fixed size~\cite{chen2013model}.

Numerous reservoir representations have been proposed~\cite{pascanu2015malware,appeltant2011information,cabessa2021efficient,chen2013model,bianchi2020reservoir}.
Currently, DPRR is considered to be the best in terms of both accuracy and circuit size~\cite{ikeda2022hardware}.
Let the elements of $\boldsymbol{x}(k)$ (i.e., the features) be $[x(k)_1, x(k)_2, \ldots, x(k)_{N_x}]$.
The vector of a node in the reservoir:
\begin{align}
 \boldsymbol{x}_n\equiv[x(1)_n, x(2)_n, \ldots, x(T)_n],
\end{align}
can be considered to be the time evolution of the node state.
Here, $T$ denotes the length of a time series.
Taking the dot product of these vectors for two nodes yields:
\begin{align}
 \boldsymbol{x}_i\cdot\boldsymbol{x}_j^- = \sum_{k=1}^{T}x(k)_ix(k-1)_j \;(i, j=1, 2, \ldots, N_x).
 \label{eq:dprr1}
\end{align}
In addition, the reservoir state itself is included as a feature:
\begin{align}
 \boldsymbol{x}_i\cdot\textbf{1} = \sum_{k=1}^{T}x(k)_i \;(i=1, 2, \ldots, N_x).
 \label{eq:dprr2}
\end{align} 
The DPRR is defined as a vector $\boldsymbol{r}$ by concatenating Eqs.~(\ref{eq:dprr1}) and (\ref{eq:dprr2}), which consists of $N_x\cdot (N_x+1)$ values, i.e.,
$\boldsymbol{r} \equiv \mbox{vec}\left(\sum^T_{i=1} \boldsymbol{x}(k) \boldsymbol{x'}(k-1)\right) \equiv [r_1, r_2, \ldots]$, where $\boldsymbol{x'}(k-1) = [\boldsymbol{x}(k), 1]$.
The output is obtained using the output layer $W_\mathrm{out}$, where $N_r$ represents the number of features in a reservoir representation:
\begin{align}
 \boldsymbol{y}=W_{\mathrm{out}}\boldsymbol{r}+\boldsymbol{b}=\sum_{n=1}^{N_r}\boldsymbol{w}_nr_n+\boldsymbol{b}.
 \label{eq:output}
\end{align}
For each of the components of $\boldsymbol{r}$, multiple reservoir states are used in the derivation, making backpropagation complicated and challenging.

\subsection{Modular DFR}\label{subsec:modular}

The challenge in designing a DFR lies in determining the design of the NL\@.
Existing analog implementations utilize elements that obey a delay differential equation, while
digital implementations that approximate the behavior of analog NL elements.

\begin{figure}[tb]
 \centering
 \includegraphics[width=0.8\linewidth]{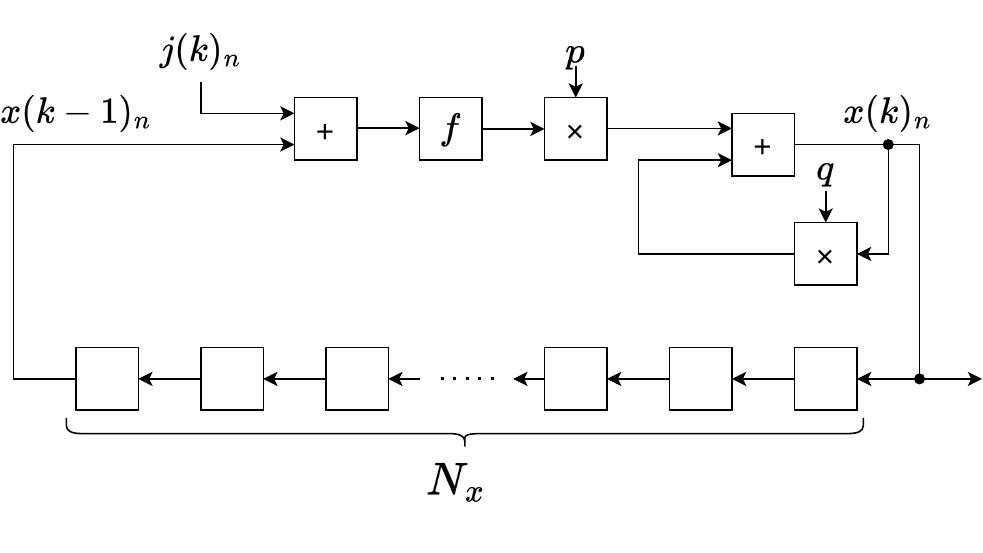}
 \caption{
 Block diagram of reservoir processing in the modular DFR model.
 The block labeled ``$f$'' operates as a one-input, one-output function $f$ in Eq.~(\ref{eq:MG}).
 Only two parameters, $p$ and $q$, have to be optimized.
 }\label{fig:model_new}
 \vspace{-5mm}
\end{figure}

The modular DFR presents a model that simplifies the design of the NL in the digital implementation of DFR~\cite{ikeda2023modular}.
To streamline the design process, the NL operation is decomposed into multiple blocks, replacing the delay differential equation with a one-input, one-output function $f$.
This simplification also reduces the number of adjustable parameters
while preserving the original solution space~\cite{ikeda2023modular}.
Fig.~\ref{fig:model_new} illustrates the block diagram of the modular DFR model.
Here, $x(k)_n$ can be represented as follows:
\begin{align}
x(k)_n=p f(j(k)_n+x(k-1)_n)+qx(k)_{n-1}.
\label{eq:modular}
\end{align}
In Eq.~(\ref{eq:modular}), the parameters to be optimized are $p$ and $q$.
Together with the regularization parameter used for linear regression in the output layer, a total of three parameters require optimization.
However, even after reducing the number of parameters, grid search remains challenging due to the need to explore the three-dimensional parameter space.
Suboptimal selection of these parameters can lead to insufficient accuracy.
Thus, an efficient optimization strategy is crucial for effective parameter tuning.

\subsection{Training of the output layer using Gaussian elimination}\label{subsec:gaussian}

\begin{figure}
\begin{algorithm}[H]
  \caption{Ridge regression using Gaussian elimination}\label{alg:gauss}
  \begin{algorithmic}[1]
  \renewcommand{\algorithmicrequire}{\textbf{Input:}}
  \renewcommand{\algorithmicensure}{\textbf{Output:}}
  \REQUIRE $A[N_y][s]$, $B[s][s]$, $B^{-1}[s][s]$, $buf$
  \ENSURE $\tilde{W}_\mathrm{out}[N_y][s]$
  \FOR {$i=0$ to $s-1$}
  \FOR {$j=0$ to $s-1$}
  \IF {$i==j$}
  \STATE $B^{-1}[i][j] \leftarrow 1$
  \ELSE
  \STATE $B^{-1}[i][j] \leftarrow 0$
  \ENDIF
  \ENDFOR
  \ENDFOR
  \FOR {$i=0$ to $s-1$}
  \STATE $buf \leftarrow 1/B[i][i]$
  \FOR {$j=0$ to $s-1$}
  \STATE $B[i][j] \leftarrow B[i][j]*buf$
  \STATE $B^{-1}[i][j] \leftarrow B^{-1}[i][j]*buf$
  \ENDFOR
  \FOR {$j=0$ to $s-1$}
  \IF {$i!=j$}
  \STATE $buf \leftarrow B[j][i]$
  \FOR {$k=0$ to $s-1$}
  \STATE $B[j][k] \leftarrow B[j][k]-B[i][k]*buf$
  \STATE $B^{-1}[j][k] \leftarrow B^{-1}[j][k]-B^{-1}[i][k]*buf$
  \ENDFOR
  \ENDIF
  \ENDFOR
  \ENDFOR
  \FOR {$i=0$ to $N_y-1$}
  \FOR {$j=0$ to $s-1$}
  \STATE $\tilde{W}_\mathrm{out}[i][j] \leftarrow 0$
  \FOR {$k=0$ to $s-1$}
  \STATE $\tilde{W}_\mathrm{out}[i][j] \leftarrow \tilde{W}_\mathrm{out}[i][j]+A[i][k]*B^{-1}[k][j]$
  \ENDFOR
  \ENDFOR
  \ENDFOR
  \end{algorithmic}
\end{algorithm}
\end{figure}

The output layer of the DFR is typically trained using Ridge regression.
The output $\boldsymbol{y}$ and the target output $\boldsymbol{e}$ are the following vectors:
\begin{align}
\boldsymbol{y}=[y_1,y_2,\ldots,y_{N_y}]\ \mathrm{and}\ \boldsymbol{e}=[e_1,e_2,\ldots,e_{N_y}].
\end{align}
Here, $N_y$ represents the number of classes, and $\boldsymbol{e}$ is represented using a one-hot encoding.
We define the following:
\begin{equation}
 \boldsymbol{\tilde{r}}\equiv[\boldsymbol{r},1].
\end{equation}
Eq.~(\ref{eq:output}) is expressed as follows:
\begin{equation}
 \boldsymbol{y}=\tilde{W}_{\mathrm{out}}\boldsymbol{\tilde{r}}.
\end{equation}
We subsequently define the following:
\begin{align}
 \tilde{R}&\equiv [\boldsymbol{\tilde{r}}^1,\boldsymbol{\tilde{r}}^2,\ldots,\boldsymbol{\tilde{r}}^{\mathrm{Train}}]
\mathrm{\ and\ }
 E \equiv [\boldsymbol{e}^1,\boldsymbol{e}^2,\ldots,\boldsymbol{e}^{\mathrm{Train}}],
\end{align}
where $\mathrm{Train}$ is the number of training data.
In Ridge regression, $\tilde{W}_{\mathrm{out}}$ is derived as follows:
\begin{align}
 \tilde{W}_{\mathrm{out}}=
  E \tilde{R}^{\mathrm{T}} {(\tilde{R}\tilde{R}^{\mathrm{T}}+\beta I)}^{-1},
 \label{eq:ridge}
\end{align}
where $I$ is an identity matrix and $\beta$ is a regularization parameter.
To simplify the following explanation, we define:
\begin{align}
 s&\equiv N_x^2+N_x+1,\\
 A&\equiv E\tilde{R}^T \in\mathbb{R}^{N_y\times s},\\
 B&\equiv \tilde{R}\tilde{R}^{\mathrm{T}}+\beta I \in\mathbb{R}^{s\times s}.
\end{align}
Then, Eq.~(\ref{eq:ridge}) is expressed as follows:
\begin{equation}
 \tilde{W}_{\mathrm{out}}=AB^{-1}.
 \label{eq:ridge2}
\end{equation}

Algorithm~\ref{alg:gauss} shows a pseudo code to perform Ridge regression by computing the matrix inversion using Gaussian elimination~\cite{arias2011suitable}.
Note that the array indexes start from zero, and that the first and second dimensions represent rows and columns, respectively.

\section{Online edge training and inference system of delayed feedback reservoir}\label{sec:propose}

\subsection{System overview}\label{subsec:overview}

In this section, we propose an online edge training and inference system of DFR characterized by the following key features.

\begin{itemize}
 \item Leveraging the modular DFR as its foundational structure, our system maintains task-specific extensibility of the NL\@.
We propose backpropagation for reservoir parameter optimization, enabling
tunable reservoir parameters that determine DFR accuracy.
This approach significantly accelerates parameter optimization and reduces training time compared to conventional systems that rely on offline grid search.
Special consideration are made for the DPRR and reservoir layers to facilitate backpropagation.
The DPRR layer, having multiple roots for a single value, necessitates backpropagation from each root.
Meanwhile, the recursive structure of the reservoir layer demands a considerable number of terms after expansion.
To enable a more lightweight implementation, we propose a truncation approach to backpropagation.

 \item To minimize the memory footprint, we propose a new matrix manipulation technique based on
Ridge regression in Eq.~(\ref{eq:ridge2}) for the output layer.
The traditional Gaussian elimination methods for computing the inverse matrix demand substantial amount of memory,
which can impede the edge implementation of output layer optimization.
Instead, our approach employs in-place Cholesky decomposition, leveraging matrix symmetry to
store approximately half the original matrix elements as a one-dimensional (1-D) array.
This technique also enables memory sharing between the original matrix
and intermediate calculations, further optimizing memory usage during matrix inversion and multiplication.
\end{itemize}

These new contributions enable the edge implementation of DFR for online training and inference, offering efficiency in terms of memory usage and runtime.
The following subsections explain the backpropagation process in its natural order for error propagation,
starting from the output layer, then moving to the DPRR layer and finally reaching the reservoir layer.

\subsection{Backpropagation in the output layer}\label{subsec:bp_output}

\begin{figure}[tb]
 \centering
 \includegraphics[width=0.7\linewidth]{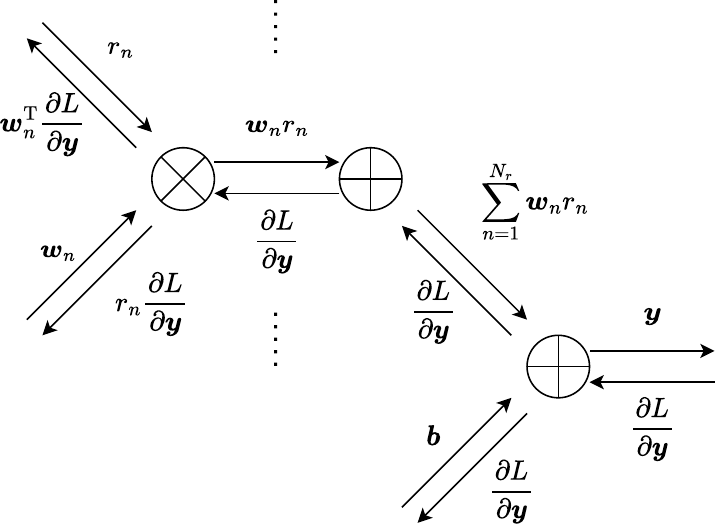}
 \caption{Computation graph of forward and backward propagation in the output layer.}\label{fig:bp_output}
\end{figure}

Fig.~\ref{fig:bp_output} illustrates the computation graph summarizing the backpropagation in the output layer.
The loss function is formulated using cross entropy error as:
\begin{equation}
L=-\sum_{k=1}^{N_y}e_k\log y_k.
\end{equation}
The partial derivative of $L$ with respect to $\boldsymbol{y}$ is given by:
\begin{equation}
\frac{\partial L}{\partial \boldsymbol{y}}=\boldsymbol{y}-\boldsymbol{e},
\end{equation}
which is propagated backwards through the network.

The computation in the output layer is expressed as Eq.~(\ref{eq:output}).
The partial derivatives of $L$ with respect to $\boldsymbol{b}$, $\boldsymbol{r}_n$, and $\boldsymbol{w}_n$ are respectively expressed as follows:
\begin{align}
\frac{\partial L}{\partial \boldsymbol{b}}=\frac{\partial L}{\partial \boldsymbol{y}}, \,
\frac{\partial L}{\partial \boldsymbol{r}_n}=\boldsymbol{w}_n^\mathrm{T}\frac{\partial L}{\partial \boldsymbol{y}}, \,
 \mathrm{and} \,
\frac{\partial L}{\partial \boldsymbol{w}_n}=\boldsymbol{r}_n\frac{\partial L}{\partial \boldsymbol{y}}.
\end{align}

\subsection{Backpropagation in the DPRR layer}\label{subsec:bp_dprr}

\begin{figure}[tbh]
 \centering
 \includegraphics[width=0.9\linewidth]{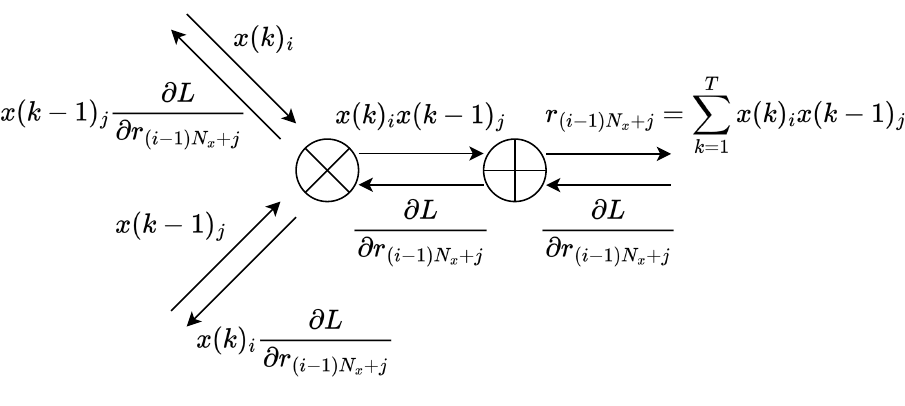}
 \caption{Computation graph of forward and backward propagation in the DPRR layer.
 }\label{fig:bp_dprr}
\end{figure}

Fig.~\ref{fig:bp_dprr} illustrates the computation graph summarizing the backpropagation in the DPRR layer.
First, each feature in the DPRR is defined as follows:
\begin{align}
r_{(i-1)N_x+j}&=\sum_{k=1}^{T}x(k)_ix(k-1)_j \;(i, j=1, \ldots, N_x),\\
r_{N_x^2+i}&=\sum_{k=1}^{T}x(k)_i \;(i=1, \ldots, N_x).
\end{align}
Errors are backpropagated from multiple reservoir representations for a feature $x(k)_n$, and the sum is backpropagated to the upstream layer.
The following backpropagated value (bpv) is propagated from the DPRR layer for the partial derivative of $L$ with respect to $x(k)_n$.
\begin{align}
(\mathrm{bpv})
 &=\sum_{j=1}^{N_x}x(k-1)_j\frac{\partial L}{\partial r_{(n-1)N_x+j}}\nonumber\\
& \quad +\sum_{i=1}^{N_x}x(k+1)_i\frac{\partial L}{\partial r_{(i-1)N_x+n}}+\frac{\partial L}{\partial r_{N_x^2+n}}.\label{eq:bpval}
\end{align}

\subsection{Backpropagation in the reservoir layer}\label{subsec:bp_reservoir}

\begin{figure}[tbh]
 \centering
 \includegraphics[width=1.0\linewidth]{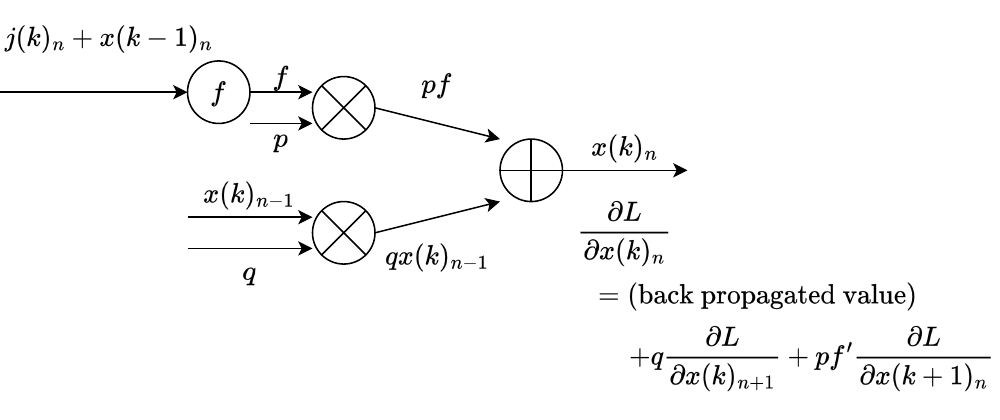}
 \caption{Computation graph of forward and backward propagation in the reservoir layer.
 }\label{fig:bp_reservoir}

\end{figure}

Fig.~\ref{fig:bp_reservoir} illustrates the computation graph summarizing the backpropagation in the reservoir layer.
Unfolding the recurrent structure of the reservoir using the modular DFR model,
the sum of errors from different nodes and different times,
and errors propagated from the downstream DPRR layer are backpropagated.
The partial derivative of $L$ with respect to $x(k)_n$ is recursively given as:
\begin{align}
\frac{\partial L}{\partial x(k)_n}&=(\mathrm{bpv})
+q\frac{\partial L}{\partial x(k)_{n+1}}+pf'\frac{\partial L}{\partial x(k+1)_n}.\label{eq:Lx}
\end{align}
The partial derivative of $L$ with respect to the constant multiplication parameters $p$ and $q$ are respectively given by summing up for all past times as follows:
\begin{align}
\frac{\partial L}{\partial p}&=\sum_{k=1}^{T}f(j(k)_n+x(k-1)_n)\frac{\partial L}{\partial x(k)_n},\label{eq:LA}\\
\frac{\partial L}{\partial q}&=\sum_{k=1}^{T}x(k)_{n-1}\frac{\partial L}{\partial x(k)_n}.\label{eq:LB}
\end{align}

\subsection{Truncated Backpropagation}\label{subsec:bp_simplified}

For $(\mathrm{bpv})$ in Eq.~(\ref{eq:bpval}), the value of $\frac{\partial L}{\partial \boldsymbol{r}}$ is determined at time $T$.
This means that the reservoir state must be stored for the number of times used for backpropagation plus one time.
Therefore, to use the entire time series, $(T+1)$ reservoir states must be stored, which consumes quadratically larger memory space as $T$ increases.
To alleviate the memory usage while maintaining optimization accuracy, we propose a \textit{truncated backpropagation} method that uses
only a limited number of reservoir states. For instance, only the last time is used.
This approximation is based on the property that the last reservoir state encompasses past reservoir states cumulatively,
with the impact of past states gradually diminishing.
With this approximation, Eqs.~(\ref{eq:bpval}),(\ref{eq:Lx}),(\ref{eq:LA}), and (\ref{eq:LB}) are respectively simplified as follows:
\begin{align}
(\mathrm{bpv})&=\sum_{j=1}^{N_x}x(T-1)_j\frac{\partial L}{\partial r_{(n-1)N_x+j}}+\frac{\partial L}{\partial r_{N_x^2+n}},\\
\frac{\partial L}{\partial x(T)_n}&=(\mathrm{bpv})+q\frac{\partial L}{\partial x(T)_{n+1}},\\
\frac{\partial L}{\partial p}&=f(j(T)_n+x(T-1)_n)\frac{\partial L}{\partial x(T)_n},\label{eq:LA2}\\
\frac{\partial L}{\partial q}&=x(T)_{n-1}\frac{\partial L}{\partial x(T)_n}.\label{eq:LB2}
\end{align}
By comparing Eqs.~(\ref{eq:LA}) and (\ref{eq:LB}) with Eqs.~(\ref{eq:LA2}) and (\ref{eq:LB2}),
it becomes evident that the computational effort is significantly reduced by approximately $1/T$.
This simplification allows us to store only two reservoir states: $\boldsymbol{x}(T-1)$ and $\boldsymbol{x}(T)$.
Specifically, for datasets with a time series length $T$ greater than 100, which holds true for most datasets,
the memory requirement for the reservoir state can be reduced to less than 2\% of the original.
When considering overall memory usage for solving a three-class classification task with a time series length of $T=500$ and $N_x=30$,
memory usage can be reduced by approximately 80\%, including other weights such as output weights.

\subsection{In-place Ridge Regression via 1-D Cholesky Decomposition}\label{subsec:cholesky}

The method proposed in the previous section enables hitherto
offline parameter search for DFR to online execution.
In addition, edge devices require learning and implementing the output layer with minimal memory footprint.
In this section, we propose in-place Ridge regression via 1-D Cholesky decomposition.

In calculating Eq.~(\ref{eq:ridge2}), matrix
$B$ is symmetric because it is expressed as follows:
\begin{eqnarray}
 B&=&(\boldsymbol{\tilde{r}}^1 \cdots \boldsymbol{\tilde{r}}^\mathrm{Train})
 \left(
 \begin{array}{c}
 \boldsymbol{\tilde{r}}^{1\ \mathrm{T}} \\
 \cdots \\
 \boldsymbol{\tilde{r}}^{\mathrm{Train}\ \mathrm{T}} \\
 \end{array}
 \right)
 +\beta I,\\
 &=&\sum_{k=1}^{\mathrm{Train}}\boldsymbol{\tilde{r}}^k\boldsymbol{\tilde{r}}^{k\ \mathrm{T}}+\beta I.
\end{eqnarray}
For any nonzero real vectors $\boldsymbol{z}$, $\boldsymbol{r}\boldsymbol{r}^\mathrm{T}$ is positive definite:
\begin{equation}
\boldsymbol{z}^\mathrm{T}(\boldsymbol{\tilde{r}}\boldsymbol{\tilde{r}}^\mathrm{T})\boldsymbol{z}=\langle\boldsymbol{z}, \boldsymbol{\tilde{r}}\rangle\langle\boldsymbol{\tilde{r}}, \boldsymbol{z}\rangle>0.
\end{equation}
Therefore, $B$ is also positive definite because it is derived as the sum of positive definite matrices.
Consequently, $B$ can be decomposed using Cholesky decomposition into the lower triangular matrix $C$:
\begin{equation}
B=CC^\mathrm{T}.
\label{eq:cholesky}
\end{equation}

Since matrix $B$ is symmetric, it suffices to store only the lower triangular components.
Additionally, by optimizing the order of decomposition computation, matrix $C$ can be derived in-place, i.e.,
matrices $C$ and $B$ can share the same memory space during decomposition requiring no additional memory.
This reduction is significant because the number of elements in $B$ is proportional to the fourth power of $N_x$.

We will perform the proposed Cholesky decomposition using a 1-D array $P$ of size $s(s+1)/2$, to facilitate hardware implementation.
The matrix $P$ initially stores the components of $B$:
\begin{equation}
P[i(i+1)/2+j]=B[i][j].
\end{equation}
The components of the lower triangle are stored sequentially in the row direction.
Similarly, the decomposed matrix is stored according to Algorithm~\ref{alg:cholesky}:
\begin{equation}
P[i(i+1)/2+j] \leftarrow C[i][j].
\end{equation}

Algorithm~\ref{alg:cholesky} provides a pseudo code for the proposed in-place Cholesky decomposition.
\begin{figure}[t]
\begin{algorithm}[H]
  \caption{Cholesky decomposition in 1-D array}\label{alg:cholesky}
  \begin{algorithmic}[1]
  \renewcommand{\algorithmicrequire}{\textbf{Input:}}
  \REQUIRE $P[s(s+1)/2]$ storing $B$
  \REQUIRE $buf$
  \ENSURE $P[s(s+1)/2]$ storing $C$
  \FOR {$i=0$ to $s-1$}
  \FOR {$j=0$ to $i-1$}
  \STATE $P[i(i+1)/2+i] \leftarrow P[i(i+1)/2+i]-P[i(i+1)/2+j]*P[i(i+1)/2+j]$
  \ENDFOR
  \STATE $P[i(i+1)/2+i] \leftarrow \mathrm{sqrt}(P[i(i+1)/2+i])$
  \STATE $buf \leftarrow 1/P[i(i+1)/2+i]$
  \FOR {$j=i+1$ to $s-1$}
  \FOR {$k=0$ to $i-1$}
  \STATE $P[j(j+1)/2+i] \leftarrow P[j(j+1)/2+i]-P[i(i+1)/2+k]*P[j(j+1)/2+k]$
  \ENDFOR
  \STATE $P[j(j+1)/2+i] \leftarrow P[j(j+1)/2+i]*buf$
  \ENDFOR
  \ENDFOR
  \end{algorithmic}
\end{algorithm}
\end{figure} 
First, $C[0][0]$ is obtained as $C[0][0]=\sqrt{B[0][0]}$ (line~5, with $i=0$) and
$C[X][0]$ is calculated by $C[X][0]=B[X][0]/C[0][0]$, where $X=1, \dots, s-1$ (line~10).
Next, $C[1][1]$ is obtained as $C[1][1]=\sqrt{B[1][1]-C[1][0]^2}$ (lines~2--5, with $i=1$) and
$C[X][1]$ is calculated by $C[X][1]=(B[X][1]-C[X][0]C[X][0])/C[1][1]$, where $X=2, \dots, s-1$ (lines~7--12, with $i=1$).
Similarly, the remaining elements can be calculated by first determining the diagonal element, and then computing the off-diagonals sequentially.
It is important to note the variable dependence in these computations---to compute each left-hand side component,
all components appearing in the right-hand side are stored in the array $P$,
necessitating no additional memory space other than a few registers.

Using Eqs.~(\ref{eq:ridge2}) and (\ref{eq:cholesky}), $A$ is expressed as:
\begin{equation}
 A = (\tilde{W}_\mathrm{out}C)C^\mathrm{T} = D C^\mathrm{T},
\end{equation}
where $D\equiv\tilde{W}_\mathrm{out}C$.
We will first compute $D$ and then $\tilde{W}_\mathrm{out}$ both in-place using
the fact that $C$ is triangular.

Algorithm~\ref{alg:calc1} shows a pseudo code to compute $D=A(C^T)^{-1}$ using backward substitution.
\begin{figure}
\begin{algorithm}[H]
  \caption{Calculation of $\tilde{W}_\mathrm{out}C$
  }\label{alg:calc1}
  \begin{algorithmic}[1]
  \renewcommand{\algorithmicrequire}{\textbf{Input:}}
  \REQUIRE $Q[N_y][s]$ storing $A$
  \REQUIRE $P[s(s+1)/2]$ storing $C$
  \ENSURE $Q[N_y][s]$ storing $D$
  \FOR {$i=0$ to $N_y-1$}
  \FOR {$j=0$ to $s-1$}
  \FOR {$k=0$ to $j-1$}
  \STATE $Q[i][j] \leftarrow Q[i][j]-Q[i][k]*P[j(j+1)/2+k]$
  \ENDFOR
  \STATE $Q[i][j] \leftarrow Q[i][j]/P[j(j+1)/2+j]$
  \ENDFOR
  \ENDFOR
  \end{algorithmic}
\end{algorithm}
\end{figure}
First, $D[0][0]$ is calculated by $D[0][0]=A[0][0]/C[0][0]$ (line~6, with $(i,j)=(0,0)$) and
then $D[0][1]$ is calculated by $D[0][1]=(A[0][1]-D[0][0]C[1][0])/C[1][1]$ (lines~3--6, with $(i,j)=(0,1)$) and
$D[0][Y]$, where $Y=2, \dots, s-1$, can be calculated sequentially.
For the other rows, the values are calculated from the left column.
Once again, this calculation is conducted entirely in-place because all components appearing in the right-hand side for the same row exist in $A$ and $Q$.
Each element in $A$ is used only once, enabling the storage space to be shared with the output.

In a similar manner, we solve $\tilde{W}_\mathrm{out}=D C^{-1}$ through the in-place forward substitution
utilizing the fact that $C$ is a lower triangular matrix.
Algorithm~\ref{alg:calc2} presents pseudo code for the computation of $\tilde{W}_\mathrm{out}$.
\begin{figure}
\begin{algorithm}[H]
  \caption{Calculation of $\tilde{W}_\mathrm{out}$
  }\label{alg:calc2}
  \begin{algorithmic}[1]
  \renewcommand{\algorithmicrequire}{\textbf{Input:}}
  \REQUIRE $Q[N_y][s]$ storing $D$
  \REQUIRE $P[s(s+1)/2]$ storing $C$
  \ENSURE $Q[N_y][n]$ storing $\tilde{W}_\mathrm{out}$
  \FOR {$i=0$ to $N_y-1$}
  \FOR {$j=s-1$ to $0$}
  \FOR {$k=s-1$ to $j+1$}
  \STATE $Q[i][j] \leftarrow Q[i][j]-Q[i][k]*P[k(k+1)/2+j]$
  \ENDFOR
  \STATE $Q[i][j] \leftarrow Q[i][j]/P[j(j+1)/2+j]$
  \ENDFOR
  \ENDFOR
  \end{algorithmic}
\end{algorithm}
\end{figure}
The calculation starts with $\tilde{W}_\mathrm{out}[0][n-1]=D[0][n-1]/C[n-1][n-1]$ (line~6, with $(i,j)=(0,n-1)$).
Next, $\tilde{W}_\mathrm{out}[0][n-2]$ is calculated by $\tilde{W}_\mathrm{out}[0][n-2]=(D[0][n-2]-\tilde{W}_\mathrm{out}[0][n-1]$ (lines~3--6, with $(i,j)=(0,n-2)$) and
$\tilde{W}_\mathrm{out}[0][Y]$, where $Y=n-3, \dots, 0$, can be similarly calculated sequentially.
For the other rows, the values are calculated from the right column.

Table~\ref{tab:comp_mem} summarizes memory consumption in data words,
which is the matrix element values, typically as 32-bit floating point numbers.
\begin{table}[t]
 \centering
 \caption{Memory footprint.
 ``Naive'' and ``Proposed'' represent the calculation via Gaussian elimination and 1-D Cholesky decomposition, respectively.
 }\label{tab:comp_mem}
 \begin{tabular}{c|cc} \toprule
   & naive & proposed \\ \midrule
  memory [word] & $2s(s+N_y)+1$ & $\frac{1}{2}s(s+2N_y)+\frac{1}{2}s$ \\ \bottomrule
 \end{tabular}
\end{table}
If $N_y$ is sufficiently small compared to $s$, the proposed method reduces memory consumption to approximately one-fourth of that required by the conventional Gaussian elimination method.
Table~\ref{tab:comp_op} compares the number of arithmetic operations.
\begin{table}[t]
 \centering
 \caption{Number of arithmetic operations.
 ``Naive'' and ``Proposed'' represent the calculation via Gaussian elimination and 1-D Cholesky decomposition, respectively.
 }\label{tab:comp_op}
 \begin{tabular}{c|cc} \toprule
  op. & naive & proposed \\ \midrule
  add & $2s^2(s+\frac{1}{2}N_y)-2s^2$ & $\frac{1}{6}s^2(s+N_y)-\frac{1}{6}s-sN_y$ \\ 
  mul & $2s^2(s+\frac{1}{2}N_y)$ & $\frac{1}{6}s^2(s+N_y)+\frac{1}{2}s^2-\frac{2}{3}s-sN_y$ \\ 
  div & $s$ & $s+2sN_y$ \\ 
  sqrt & $0$ & $s$ \\ \bottomrule
 \end{tabular}
\end{table}
Under similar conditions, when $N_y$ is sufficiently small compared to $s$, the proposed method reduces the number of additions and multiplications to about one-twelfth of those required by the Gaussian elimination method.
\section{Evaluation}\label{sec:evaluation}

In this section, we evaluate the fast reservoir parameter optimization method,
the memory-saving Ridge regression,
and the online edge training and inference system achieved by combining them.
The evaluation uses the same datasets (npz files) as in~\cite{bianchi2020reservoir}.
Table~\ref{tab:dataset} provides a summary of the datasets used.
The evaluations consistently used $f(x)=\alpha x$ as the nonlinear function in the modular DFR for all datasets, as recommended in~\cite{ikeda2023modular}.
The reservoir size $N_x$ was set to 30.

\subsection{Evaluation of the proposed fast parameter optimization for the reservoir layer}\label{subsec:eval1}

The proposed parameter optimization method was evaluated
on a Windows workstation equipped with an AMD Ryzen 7 5825U CPU and 16~GB of memory.
All programs were implemented in Python 3.9.6 and utilized the numpy library version 1.23.0.
\begin{table}
 \centering
 \caption{Summary of multivariate time series classification datasets (taken from \cite{bianchi2020reservoir}).
 Column \#$V$, \#$C$, Train, and Test respectively report
 the input dimension, the output dimension, the number of training
 data, and that of testing data.  $T_\mathrm{min}$ and
 $T_\mathrm{max}$ are the shortest and the longest length of the input
 series in the dataset, respectively.}
 \label{tab:dataset}
 \begin{tabular}{c|cccccc} \toprule
  Dataset & \#$V$ & \#$C$ & Train & Test & $T_{\mathrm{min}}$ & $T_{\mathrm{max}}$\\ \midrule
  ARAB & 13 & 10 & 6600 & 2200 & 4 & 93 \\
  AUS & 22 & 95 & 1140 & 1425 & 45 & 136 \\
  CHAR & 3 & 20 & 300 & 2558 & 109 & 205 \\
  CMU & 62 & 2 & 29 & 29 & 127 & 580 \\
  ECG & 2 & 2 & 100 & 100 & 39 & 152 \\
  JPVOW & 12 & 9 & 270 & 370 & 7 & 29 \\
  KICK & 62 & 2 & 16 & 10 & 274 & 841 \\
  LIB & 2 & 15 & 180 & 180 & 45 & 45 \\
  NET & 4 & 13 & 803 & 534 & 50 & 994 \\
  UWAV & 3 & 8 & 200 & 427 & 315 & 315 \\
  WAF & 6 & 2 & 298 & 896 & 104 & 198 \\
  WALK & 62 & 2 & 28 & 16 & 128 & 1918 \\ \bottomrule
 \end{tabular}
\end{table}

In the proposed method, the reservoir parameters $p$ and $q$, along with the output-layer parameters $W_{\mathrm{out}}$ and $b$ are optimized using the proposed truncated backpropagation.
The initial values of $[p, q]$ are $[0.01,0.01]$, and the output parameters are initialized to zeros.
The stochastic gradient descent is employed for 25 epochs.
The learning rate started at 1 and was reduced by multiplying 0.1
for the reservoir layer parameters at epochs 5, 10, 15, and 20.
For the output layer parameters, the learning rate was multiplied by 0.1 at epochs 10, 15, and 20.
Once the optimization with backpropagation is completed, the output layer is trained using Ridge regression with the regularization parameter $\beta$.
We evaluate $\beta$ values of $[10^{-6}, 10^{-4}, 10^{-2}, 10^{0}]$ and choose the one resulting in the lowest loss $L$.

We compared the proposed method with grid search,
which was executed in three dimensions: $p$, $q$, and $\beta$.
The ranges for $p$ and $q$ were set to $[10^{-3.75},10^{-0.25}]$, $[10^{-2.75},10^{-0.25}]$, respectively.
These ranges were determined to cover the optimal parameters for all the datasets used in this evaluation.
The number of grid divisions is increased from 1 until the accuracy matches that of the proposed method.
Note that the grid divisions are equidistant, e.g., the range for $p$ and $q$ are simultaneously divided into sections of equal length.
$\beta$ is searched in the same way as the proposed method.

Table~\ref{tab:vsgrid} summarizes the comparison results for accuracy and computation time.
The parameters optimized with backpropagation attained equal or superior accuracy to the state-of-the-art~\cite{ikeda2022hardware} for most datasets,
while significantly reducing the computation time to approximately 1/700 of that of the grid search.
\begin{table}
 \centering
 \caption{Runtime comparison between the proposed backpropagation (bp) and grid search (gs).
 Column ``gs divs'' represents the number of grid divisions required to achieve
 equal accuracy as the proposed method (bp acc).
 }\label{tab:vsgrid}
 \begin{tabular}{c|ccccc} \toprule
   & bp & bp & gs & gs & (gs time)/ \\
  dataset & acc & time (s) & divs & time (s) & (bp time) \\ \midrule
  ARAB & 0.981 & 245 & 8 & 25,040 & 102.2 \\
  AUS & 0.954 & 54 & 8 & 5,535 & 102.5 \\
  CHAR & 0.918 & 44 & 10 & 4,820 & 109.5 \\
  CMU & 0.931 & 4 & 1 & 3 & 0.8 \\
  ECG & 0.850 & 11 & 16 & 4,977 & 452.5 \\
  JPVOW & 0.978 & 4 & 4 & 106 & 26.5 \\
  KICK & 0.800 & 7 & 1 & 2 & 0.3 \\
  LIB & 0.806 & 12 & 18 & 8,423 & 701.9 \\
  NET & 0.783 & 45 & 1 & 49 & 1.1 \\
  UWAV & 0.850 & 65 & 10 & 6,322 & 97.3 \\
  WAF & 0.983 & 14 & 3 & 188 & 13.4 \\
  WALK & 1.000 & 4 & 1 & 3 & 0.8 \\ \bottomrule
 \end{tabular}
\end{table}

\begin{figure}[tb]
 \centering
 \includegraphics[width=0.8\linewidth]{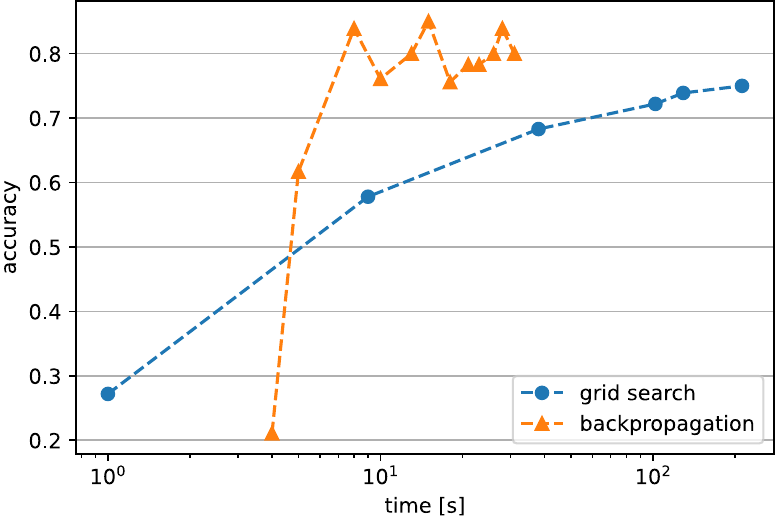}
 \caption{Accuracy versus computation time for the dataset LIB\@.
 }\label{fig:vs}
\end{figure}

Fig.~\ref{fig:vs} compares the parameter optimization times between the
proposed method and grid search. The results demonstrate that the 
proposed method not only achieves superior accuracy---closer to the true optimal parameters---but also reduces
computation time by several orders of magnitude.
The reduction in computation time is particularly significant for datasets where grid search requires a long computation time.
For datasets requiring fewer grid divisions indicate either the optimal parameters are widely spread, or that the initial coarse grid coincidentally yields satisfactory outcomes.
However, the practical challenges of grid search include the difficulty of determining whether the best accuracy has been attained, making early stopping unreliable.
In contrast, datasets with sharp and narrowly defined optimal parameters necessitate very fine grid divisions for accurate optimization.
In such cases, grid search may demand exponentially long computation time.

Table~\ref{tab:acc} compares classification accuracy with other machine learning methods.
The machine learning methods that were compared were:
\begin{enumerate}
\item Multi layer perceptron (MLP) \cite{wang2017time}
\item Fully convolutional neural network (FCN) \cite{wang2017time}
\item Residual network (ResNet) \cite{wang2017time}
\item Encoder \cite{serra2018towards}
\item Multi channel deep convolutional neural network (MCDCNN) \cite{zheng2014time}
\item Time convolutional neural network (Time-CNN) \cite{zhao2017convolutional}
\item Time warping invariant echo state network (TWIESN). \cite{tanisaro2016time}
\end{enumerate}
\begin{table*}
\centering
 \caption{Classification accuracy comparison with other machine learning methods. The data in the table, except for prop. bp, are from~\cite{IsmailFawaz2018deep}.}
\label{tab:acc}
\begin{tabular}{c|cccccccc} \toprule
dataset & MLP & FCN & ResNet & Encoder & MCDCNN & Time-ECN & TWIESN & prop. bp \\ \midrule
ARAB & 0.969 & 0.994 & 0.996 & 0.981 & 0.959 & 0.958 & 0.853 & 0.981 \\
AUS & 0.933 & 0.975 & 0.974 & 0.938 & 0.854 & 0.726 & 0.724 & 0.954 \\
CHAR & 0.969 & 0.990 & 0.990 & 0.971 & 0.938 & 0.960 & 0.920 & 0.918 \\
CMU & 0.600 & 1.000 & 0.997 & 0.983 & 0.514 & 0.976 & 0.893 & 0.931 \\
ECG & 0.748 & 0.872 & 0.867 & 0.872 & 0.500 & 0.841 & 0.737 & 0.850 \\
JPVOW & 0.976 & 0.993 & 0.992 & 0.976 & 0.944 & 0.956 & 0.965 & 0.978 \\
KICK & 0.610 & 0.540 & 0.510 & 0.610 & 0.560 & 0.620 & 0.670 & 0.800 \\
LIB & 0.780 & 0.964 & 0.954 & 0.783 & 0.651 & 0.637 & 0.794 & 0.806 \\
NET & 0.550 & 0.891 & 0.627 & 0.777 & 0.630 & 0.890 & 0.945 & 0.783 \\
UWAV & 0.901 & 0.934 & 0.926 & 0.908 & 0.845 & 0.859 & 0.754 & 0.850 \\
WAF & 0.894 & 0.982 & 0.989 & 0.986 & 0.658 & 0.948 & 0.949 & 0.983 \\
WALK & 0.700 & 1.000 & 1.000 & 1.000 & 0.450 & 1.000 & 0.944 & 1.000 \\ \bottomrule
\end{tabular}
\end{table*}
The proposed backpropagation method (prop. bp) achieves an accuracy better than or comparable to all these machine learning methods.

\begin{figure}[tb]
 \centering
 \includegraphics[width=0.8\linewidth]{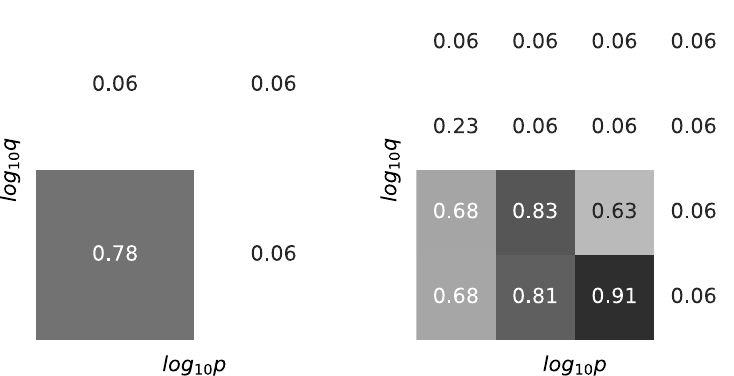}
 \caption{Grid search example where recursively subdividing the previously best grid is challenging.
 The dataset used is CHAR\@. Left: grid level 1, right: grid level 2.}\label{fig:grid}
 \vspace{-5mm}
\end{figure}

An alternative approach to explore a finer parameter space is to recursively divide the best region in the coarser search.
However, this method may result in suboptimal parameters and fail for certain datasets if the coarse grid search does not yield globally optimal parameters,
as illustrated in Fig.~\ref{fig:grid}.
The proposed method successfully identified optimal values with a fixed number of epochs for all datasets used in this study.
Although exploring different initial values or increasing epochs could potentially lead to further improvements, they did not enhance the accuracy of the presented datasets.

\begin{table}
 \centering
 \caption{Storage reduction by truncated backpropagation.
 ``Naive'' and ``simplified'' represent the total number of stored values of the reservoir state, reservoir representation, and reservoir weight,
 before and after truncating the computation of backpropagation.
The datapath is consistently calculated by a word, which is float32.
 }\label{tab:comp3}
 \begin{tabular}{c|ccc} \toprule
   & naive & simplified & (naive-simplified)/ \\
  dataset & [word] & [word] & naive \\ \midrule
  ARAB & 13,030 & 10,300 & 21~\% \\
  AUS & 93,455 & 89,435 & 4~\% \\
  CHAR & 25,700 & 19,610 & 24~\% \\
  CMU & 20,192 & 2,852 & 86~\% \\
  ECG & 7,352 & 2,852 & 61~\% \\
  JPVOW & 10,179 & 9,369 & 8~\% \\
  KICK & 28,022 & 2,852 & 90~\% \\
  LIB & 16,245 & 14,955 & 8~\% \\
  NET & 42,853 & 13,093 & 69~\% \\
  UWAV & 17,828 & 8,438 & 53~\% \\
  WAF & 8,732 & 2,852 & 67~\% \\
  WALK & 60,332 & 2,852 & 95~\% \\ \bottomrule
 \end{tabular}
\end{table}
Table~\ref{tab:comp3} shows the memory savings resulting from the truncation of the backpropagation calculations.
While the reduction in memory is relatively smaller for datasets with a large number of classes and short input series lengths,
more than half of the datasets still achieve a reduction of over 50\%.
In addition, for all datasets, even those with small memory reductions,
the computational effort of backpropagation has been reduced to approximately $1/T$,
resulting in a considerable reduction in computation time.

\subsection{Evaluation of the proposed memory-saving Ridge regression for the output layer}

Next, we evaluate the proposed in-place Ridge regression via 1-D Cholseky Decomposition.
The evaluation environment remains unchanged from Section~\ref{subsec:eval1}.
The optimal parameters obtained by the proposed parameter optimization in Section~\ref{subsec:eval1} are used to perform Ridge regression.
The naive method with Gaussian elimination and the proposed method with 1-D Cholesky decomposition are compared.

\begin{figure}[tb]
 \centering
 \includegraphics[width=0.8\linewidth]{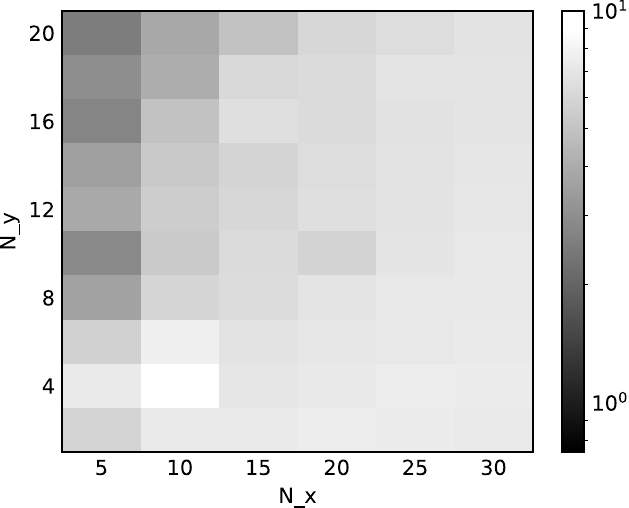}
 \caption{Runtime ratio of the calculation via Gaussian elimination to via Cholesky decomposition. 
 }\label{fig:pcolor}
\end{figure}

Fig.~\ref{fig:pcolor} plots the ratio of computation times
between Gaussian elimination and proposed Cholesky decomposition.
Lighter shades represent faster computation times achieved by the proposed method relative to the Gaussian elimination.
The proposed method demonstrates consistent and significant speed advantages for practical $N_x$ values exceeding 10.
Notably, for $N_y$ values below 10, the proposed method achieves a speedup of approximately 7x.
While the proposed method introduces additional division and square root operations,
the substantial reduction in the number of additions and multiplications significantly reduces the total computation time.

Table~\ref{tab:vschol} shows the memory savings achieved by using the proposed Ridge regression.
\begin{table}
 \centering
 \caption{Memory usage in Ridge regression. 
 ``Naive'' and``Prop.'' use Gaussian elimination and Cholesky decomposition, respectively.
 A 32-bit float is evaluated as one word.
 }\label{tab:vschol}
 \begin{tabular}{c|ccccc} \toprule
  data & \multicolumn{2}{c}{accuracy} & \multicolumn{2}{c}{memory [word]]} & (naive mem)/ \\
  set  & naive & prop.          & naive &  prop.             & (prop. mem) \\ \midrule
  ARAB & 0.981 & 0.981 & 1,752,142 & 443,156 & 3.95 \\ 
  AUS & 0.954 & 0.954 & 1,910,412 & 522,291 & 3.66 \\ 
  CHAR & 0.918 & 0.918 & 1,770,762 & 452,466 & 3.91 \\
  CMU & 0.931 & 0.931 & 1,737,246 & 435,708 & 3.99 \\
  ECG & 0.850 & 0.850 & 1,737,246 & 435,708 & 3.99 \\
  JPVOW & 0.978 & 0.978 & 1,750,280 & 442,225 & 3.96 \\
  KICK & 0.800 & 0.800 & 1,737,246 & 435,708 & 3.99 \\
  LIB & 0.806 & 0.806 & 1,761,452 & 447,811 & 3.93 \\
  NET & 0.783 & 0.783 & 1,757,728 & 445,949 & 3.94 \\
  UWAV & 0.850 & 0.850 & 1,748,418 & 441,294 & 3.96 \\
  WAF & 0.983 & 0.983 & 1,737,246 & 435,708 & 3.99 \\
  WALK & 1.000 & 1.000 & 1,737,246 & 435,708 & 3.99 \\ \bottomrule
 \end{tabular}
\end{table}
Despite achieving the same accuracy as the naive method, the proposed method reduces
memory usage by almost four times for all datasets.

\subsection{Hardware implementation}

The proposed parameter optimization method and Ridge regression
were used to implement an online edge training and inference system
of DFR on an FPGA board using Zynq-7000 (xc7z020clg400-1).
The C++ code was synthesized in Xilinx Vitis HLS 2021.1,
and the circuit was implemented and evaluated in Xilinx Vivado 2021.1.
Although this paper demonstrates
limited synthetic examples,
HLS enables flexible adjustments to trade-off between power and resource consumption, 
making it especially suitable for edge computing applications.

The hardware implementation was relatively straightforward owing to the design of the proposed algorithms.
However, bottlenecks were identified in the computation
of reservoir representations, backpropagation, and ridge regression.
To address these challenges, loops within these processes were carefully pipelined.
In particular, memory dependencies arising from successive accesses to the same address locations,
as shown in line 4 of Algorithm~\ref{alg:calc1}, constrained the reduction of iteration intervals.
To mitigate this limitation, small write buffers were introduced. Consequently,
Algorithm~\ref{alg:calc1} was revised into Algorithm~\ref{alg:calc1imp},
where $Q$ was partitioned to enable parallelism.
Similar optimization was applied to Algorithm~\ref{alg:calc2}.

\begin{figure}
\begin{algorithm}[H]
  \caption{Improved calculation of $\tilde{W}_\mathrm{out}C$}
  \label{alg:calc1imp}
  \begin{algorithmic}[1]
  \renewcommand{\algorithmicrequire}{\textbf{Input:}}
  \REQUIRE $Q[N_y][s]$ storing $A$
  \REQUIRE $P[s(s+1)/2]$ storing $C$
  \REQUIRE $reg[\mathrm{RegSize}+1]$
  \ENSURE $Q[N_y][s]$ storing $D$
  \FOR {$i=0$ to $N_y-1$}
  \FOR {$j=0$ to $s-1$}
  \FOR {$k=0$ to $j-1$}
  \STATE $\mathrm{mul} \leftarrow Q[i][k]*P[j(j+1)/2+k]$
  \STATE $\mathrm{idx} \leftarrow 0$
  \FOR {$l=0$ to $\mathrm{RegSize}-1$}
  \IF {$l==0$}
  \IF{$\mathrm{idx}<\mathrm{RegSize}$}
  \STATE $\mathrm{reg}[\mathrm{RegSize}] \leftarrow -\mathrm{mul}$
  \ELSE
  \STATE $\mathrm{reg}[\mathrm{RegSize}] \leftarrow \mathrm{reg}[0]-\mathrm{mul}$
  \ENDIF
  \ENDIF
  \STATE $\mathrm{reg}[l] \leftarrow \mathrm{reg}[l+1]$
  \ENDFOR
  \STATE $\mathrm{idx} \leftarrow \mathrm{idx}+1$
  \ENDFOR
  \FOR {$k=0$ to $\mathrm{RegSize}$}
  \STATE $Q[i][j] \leftarrow Q[i][j]-reg[k]$
  \ENDFOR
  \STATE $Q[i][j] \leftarrow Q[i][j]/P[j(j+1)/2+j]$
  \ENDFOR
  \ENDFOR
  \end{algorithmic}
\end{algorithm}
\end{figure}
\begin{figure}[tb]
 \centering
 \includegraphics[width=0.8\linewidth]{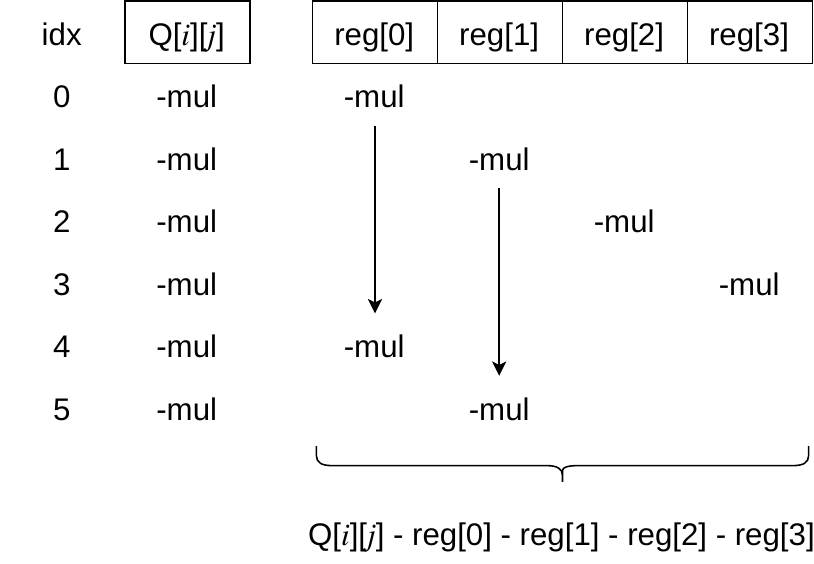}
 \caption{Conceptual timing diagram showing the efficacy of adding a small write buffer.}\label{fig:reg}
\end{figure}

Fig.~\ref{fig:reg} illustrates the effectiveness of incorporating the write buffer.
In the naive implementation, multiplication and subtraction operations followed by memory writes had to complete within a single clock period,
significantly limiting the achievable clock frequency. In contrast,
the improved implementation allows these operations to finish before the subsequent loop iteration after all write buffers are filled.
This buffer significantly relaxes timing constraints and enabling pipelining at the cost of increased DSP usage.
The value of $\mathrm{RegSize}$ was carefully determined to balance trade-offs.
A larger
$\mathrm{RegSize}$ allows for greater parallel processing in lines 3--17 of Algorithm~\ref{alg:calc1imp}, but it also increases
loop iterations in lines 18--20 and introduces memory access conflicts.
After balancing these considerations, the value of $\mathrm{RegSize}$ was set to 4 throughout this work.

Initially, prior to applying these optimizations,
bottlenecks were identified in three key modules: reservoir representation, back propagation, and ridge regression.
Among these, ridge regression posed the most significant bottleneck due to deeply nested memory-conflicting loops.
The proposed optimizations resolved these conflicts, eliminating the bottleneck in ridge regression and shifting it to
the reservoir state update, represented by Eqs.~(\ref{eq:digitalDFR1}) and (\ref{eq:digitalDFR2}), which now determines the maximum achievable clock frequency.

To evaluate performance,
a comparison was conducted between two implementation approaches:
executing the entire process on the FPGA (HW-only) and running
the corresponding C++ code using the processor on the same evaluation board (SW-only).
The JPVOW dataset was used for the following evaluations.

Table~\ref{tab:hard} compares the implementation results.
\begin{table}
 \centering
 \caption{Performance comparison between fully hardware (FPGA) and fully software implementations
 }\label{tab:hard}
 \begin{tabular}{c|cc} \toprule
   & SW only & HW only \\ \midrule
  LUT & - & 33,674 (63.2\%) \\ 
  LUTRAM & - & 1,073 (6.2\%) \\
  FF & - & 49,596 (46.6\%) \\
  BRAM & - & 26.50 (18.9\%) \\
  DSP & - & 143 (65.0\%) \\
  BUFG & - & 1 (3.1\%) \\
  Clock frequency & 667\,MHz & 100\,MHz \\
  FPGA power consumption & - & 0.734\,W \\
  Processor power consumption & 1.530\,W & -  \\  
  Calculation time & 5.56\,s & 0.42\,s \\
  Training time & 5.42\,s & 0.32\,s \\
  Inference time & 0.14\,s & 0.10\,s \\
  Energy consumption & 8.51\,J & 0.31\,J \\ \bottomrule
 \end{tabular}
\end{table}
In hardware resource utilization, numbers in parentheses indicate the percentage of available FPGA resources used.
The BRAM size is 36\,kb per block.
Power consumption was estimated using the Vivado tool after the place and route process,
and the ``Calculation time'' represents the combined duration of training and inference executed sequentially.
While the accuracy achieved by the software and hardware implementations is identical,
the hardware implementation operates at a significantly lower maximum clock frequency compared to
the Dual-core ARM Cortex-A9 MPCore processor on the FPGA board. Despite this, the
hardware implementation reduced computation time to approximately 1/13 and power consumption to about 1/27.
These results demonstrate the suitability of the proposed DFR system for execution on edge-hardware.
Table~\ref{tab:module} details resource utilization for each major module.
Here, the DFR core module is responsible for processing the signal as it passes through the input, reservoir, and output layers.
Note that these values were estimated after HLS and do not account for the resources used during the place and route process.
\begin{table}
 \centering
 \caption{Resource utilization of major modules
 }\label{tab:module}
 \begin{tabular}{c|ccc} \toprule
   & DFR core & backpropagation & ridge regression  \\ \midrule
  LUT & 8,764 & 12,245 & 7,827  \\ 
  FF & 11,266 & 10,125 & 8,228 \\
  DSP & 15 & 57 & 20 \\ \bottomrule
 \end{tabular}
\end{table}

Table~\ref{tab:pragma} presents synthesis results under different constraints, comparing
non-pipelined and inlined configurations. The
non-pipelined configuration was intended to minimize circuit size without incorporating pipelining, while the
inlined configuration expanded the reservoir state update process inline, addressing a bottleneck in the standard implementation.
These results illustrate the Pareto front of the implementations, highlighting the flexibility offered by HLS for adopting various scenarios typical in edge computing environments.
\begin{table}
 \centering
 \caption{Performance comparison of non-pipelined and inlined configurations
 }\label{tab:pragma}
 \begin{tabular}{c|cc} \toprule
   & non-pipelined &  inlined \\ \midrule
  LUT & 22,680 (42.6\%) & 44,237 (83.2\%) \\ 
  LUTRAM & 755 (4.3\%) & 884 (5.1\%) \\
  FF & 31,953 (30.0\%) & 59,726 (56.13\%) \\
  BRAM & 25.50 (18.2\%) & 27.50 (19.64\%) \\
  DSP & 121 (55.0\%) & 136 (61.82~\%) \\
  BUFG & 1 (3.1\%) & 1 (3.1\%) \\
  Clock frequency & 100\,MHz & 100\,MHz \\
  FPGA power consumption & 0.704\,W & 0.864\,W \\
  Calculation time & 1.44\,s & 0.38\,s \\
  Training time & 1.36\,s & 0.29\,s \\
  Inference time & 0.09\,s & 0.09\,s \\
  Energy consumption & 1.01\,J & 0.33\,J \\ \bottomrule
 \end{tabular}
\end{table}

\begin{table}
 \centering
 \caption{Comparison with existing FPGA implementations of DFR\@.
 Columns \#$V$ and \#$C$ respectively report the input and output dimensions.}
 \label{tab:dataset2}
 \begin{tabular}{c|cccccc} \toprule
  method & training/inference on HW & implementation & \#$V$ & \#$C$\\ \midrule
  prop. & both & fully digital & 12 & 9 \\
  \cite{alomar2015digital} & inference only & fully digital & 1 & 3 \\
  \cite{shears2021hybrid} & inference only & digital/analog hybrid & 1 & 1 \\ \bottomrule
 \end{tabular}
\end{table}
Table~\ref{tab:dataset2} compares the proposed FPGA implementation of DFR with existing approaches.
Unlike previous methods, which are limited to low-dimensional input-output tasks that are relatively simple to implement,
the proposed method supports practical and effective applications by performing both training and inference for multidimensional input-output tasks directly and entirely on the FPGA\@.

%

\section{Conclusion}\label{sec:conclusion}

We proposed an online edge training and inference system for DFR, introducing several key innovations.
First, we proposed a fast and accurate parameter optimization method for the reservoir layer using backpropagation and gradient descent.
By adopting the modular DFR model, we formulated backpropagation for DPRR, 
enhancing accuracy while addressing computational challenges caused by long state dependencies.
Additionally, a truncated backpropagation method was introduced, leveraging reservoir state properties to reduce computation time 
by up to approximately 1/700 and halve memory usage for most datasets.
For the output layer, We proposed an in-place Ridge regression based on 1-D Cholesky decomposition.
By utilizing matrix symmetry and optimizing the array update order, this method reduced memory usage by approximately 75\% while maintaining accuracy.
Finally, we implemented the entire system on an FPGA, incorporating these methods.
The hardware implementation achieved an 13x reduction in computation time and a 27x reduction in
power consumption simultaneously, compared to the software implementation on a dual-core ARM Cortex-A9 processor.

\bibliographystyle{ACM-Reference-Format}
\bibliography{ref}

\end{document}